\documentclass[graybox]{svmult}
\usepackage[pdftex]{graphicx}
\usepackage{type1cm}        % activate if the above 3 fonts are
\usepackage{makeidx}         % allows index generation
\usepackage{multicol}        % used for the two-column index
\usepackage[bottom]{footmisc}% places footnotes at page bottom
\usepackage{cite}

\usepackage{newtxtext}       % 
\usepackage[varvw]{newtxmath}       % selects Times Roman as basic font

\def\[#1\]{\begin{align}#1\end{align}}
\makeindex             % used for the subject index
\begin{document}

\title*{CDT and Ho\v rava-Lifshitz QG\\ in Two Dimensions}
% Use \titlerunning{Short Title} for an abbreviated version of
% your contribution title if the original one is too long
\author{Yuki Sato}
% Use \authorrunning{Short Title} for an abbreviated version of
% your contribution title if the original one is too long
\institute{Yuki Sato \at National Institute of Technology, Tokuyama College, Gakuendai, Shunan, Yamaguchi 745-8585, Japan, \email{sato@tokuyama.ac.jp}
\and 
%Name of Second Author \at 
Nagoya University, Chikusaku, Nagoya 464-8602, Japan, \email{ysato.phys@gmail.com}}
%
% Use the package "url.sty" to avoid
% problems with special characters
% used in your e-mail or web address
%
\maketitle

\abstract*{
%Each chapter should be preceded by an abstract (no more than 200 words) that summarizes the content. The abstract will appear \textit{online} at \url{www.SpringerLink.com} and be available with unrestricted access. This allows unregistered users to read the abstract as a teaser for the complete chapter.\newline\indent
%Please use the 'starred' version of the \texttt{abstract} command for typesetting the text of the online abstracts (cf. source file of this chapter template \texttt{abstract}) and include them with the source files of your manuscript. Use the plain \texttt{abstract} command if the abstract is also to appear in the printed version of the book.
}

\abstract{
Two-dimensional causal dynamical triangulations ($2$d CDT) is a lattice model of quantum geometry. 
In $2$d CDT, one can deal with the quantum effects analytically and explore the physics through the continuum limit. 
The continuum theory is known to be two-dimensional projectable Ho\v rava-Lifshitz quantum gravity ($2$d projectable HL QG). 
In this chapter, we wish to review the very relation between $2$d CDT and $2$d projectable HL QG in detail.        
}

\vspace{20pt}
\noindent \textbf{Keywords}\\
\\
Lattice quantum gravity; Causal dynamical triangulations (CDT); Ho\v rava-Lifshitz quantum gravity; Two-dimensional models; Wormholes

\vspace{20pt}
\noindent \textbf{Note}
This is a preprint of the chapter in the Handbook of Quantum Gravity. 
It belongs to the section 
entitled Causal Dynamical Triangulations.

\pagebreak

\section{Introduction}
\label{sec:introduction}

Two-dimensional toy models of quantum gravity are a very useful playground for understanding the quantum nature of geometry quantitatively.   
This is because many of them are simple enough to be dealt with analytically and complex enough to observe nontrivial quantum effects. 
Lattice regularizations in particular are known to be quite powerful tools for investigating non-perturbative quantum effects analytically. 
Among these, two-dimensional Euclidean dynamical triangulations ($2$d EDT) \cite{Ambjorn:1985az,Ambjorn:1985dn,David:1984tx,Billoire:1985ur,Kazakov:1985ea,Boulatov:1986jd} 
(see a pedagogical textbook \cite{Ambjorn:1997di}) and two-dimensional causal dynamical triangulations ($2$d CDT) \cite{Ambjorn:1998xu} (see a detailed review \cite{Ambjorn:2022btk}) are good practical examples. 
The former and the latter, respectively, are Euclidean and Lorentzian lattice models 
based on Regge's discretization of geometries \cite{Regge:1961px}. 

$2$d EDT discretizes Euclidean geometries by equilateral triangles and defines a regularized quantum amplitude as a sum over distinct triangulated geometries. 
Matrix models and combinatorics can be used for calculating such a statistical sum analytically (whenever possible). 
By virtue of analytic treatments, one can explicitly remove the regularization through the continuum limit to calculate physical observables. 
What is remarkable is that exactly the same value of observables can be reproduced from a genuine continuum field theory called the Liouville quantum gravity \cite{Polyakov:1981rd, Knizhnik:1988ak, David:1988hj, Distler:1988jt}. 
This means that $2$d EDT serves as a well-defined regularization of the Liouville quantum gravity.

$2$d CDT is a Lorentzian lattice model of quantum geometry which respects a global time foliation 
and prohibits the creation of so-called baby universes. 
One can calculate the sum over such Lorentzian triangulated geometries using simple combinatorics, 
and take the continuum limit to remove the regularization. 
All these processes can be performed analytically at least for the plain model without coupling to a matter. 
It has been shown in Ref.~\cite{Ambjorn:2013joa} that the resulting continuum theory is known to be in the same universality class of projectable Ho\v rava-Lifshitz quantum gravity (projectable HL QG) \cite{horava} in two dimensions, which is different from the Liouville quantum gravity \footnote{In fact, it has been shown that a direct lattice discretization of $2$d projectable HL gravity, 
which has a lattice action different from that of $2$d CDT, reproduces the same large-scale physics in the continuum limit \cite{Glaser:2016smx}.}.

$2$d HL QG is a quantum field theory in two dimensions, 
which has a preferred foliation structure. 
This model is invariant only under the subclass of diffeomorphisms that respects the foliation, 
known as the foliation-preserving diffeomorphisms\footnote{At the cost of full diffeomorphism invariance,  
HL QG has been designed originally as a model of quantum gravity in higher dimensions such that it has a good convergence at UV in keeping with unitarity, 
and would approximately recover the diffeomorphism invariance at IR \cite{horava}}. 
$2$d projectable HL QG is a certain version of HL QG where a part of the time-time component of the metric called the lapse function is projectable, i.e. a function only of time. 
In this chapter, we wish to explain in detail the relation between projectable HL QG and  CDT in two dimensions\footnote{The relation between HL QG and CDT in four dimension has been pointed out first in Ref.~\cite{Horava:2009if} by looking at an observable called the spectral dimension, and the resemblance of the CDT phase diagram to a Lifshitz phase diagram has been shown in Ref.~\cite{Ambjorn:2010hu}.}.

In fact one can generalize $2$d CDT in such a way that the creation and annihilation of (a finite number of) baby universes, and the formation of wormholes (handles) are allowed to occur in keeping with the foliation structure. 
This model is called the generalized CDT (GCDT) introduced first as a continuum theory \cite{Ambjorn:2007jm, Ambjorn:2008ta} and later defined at the discrete level \cite{Ambjorn:2008gk, Ambjorn:2013csx}. 
The full continuum description of GCDT is given by the so-called string field theory for CDT \cite{Ambjorn:2008ta} in which the string means the one-dimensional closed spatial universe, 
and the baby universes and wormholes can be realized in terms of the splitting and joining interactions of strings.

One of remarkable facts is that focusing on a certain amplitude, i.e. loop-to-loop amplitude, one can read off a one-dimensional effective theory 
that takes in all possible baby universe and wormhole contributions in an effective manner \cite{Ambjorn:2009wi, Ambjorn:2009fm}: 
The $1$d effective theory is a one-body quantum theory even though GCDT is a many-body theory that allows both creation and annihilation of strings. 
It is known that one can correctly reproduce the $1$d effective theory if quantizing the projectable HL gravity with a certain bi-local interaction term \cite{Ambjorn:2021wou, Ambjorn:2021ysb}. 
This topic will be treated in this chapter.

Furthermore, the $1$d effective theory that includes all contributions of baby universes and wormholes is known to be reproduced 
if one assumes that the cosmological constant of the continuum limit of $2$d CDT is not really a constant but fluctuates in time \cite{Ambjorn:2021wdm}. 
This idea leads to a certain realization of Coleman's mechanism about the cosmological constant \cite{Coleman:1988tj} in the context of CDT \cite{Ambjorn:2021wdm}, 
which will be also explained in this chapter.

The rest of this chapter is organized as follows. 
In Sec.~\ref{sec:2dcdt}, a self-contained introduction to $2$d CDT is presented. 
We show that taking the continuum limit the physics of $2$d CDT can be described as a one-dimensional quantum system with a Hamiltonian.    
$2$d projectable HL QG is explained in Sec.~\ref{sec:2dHLQG}. Through the path-integral quantization we read off the quantum Hamiltonian that is equivalent to the one obtained in the continuum limit of $2$d CDT. 
Thereby one can confirm that the continuum limit of $2$d CDT is $2$d projectable HL QG.   
In Sec.~\ref{sec:sumoverallgeneraingenera}, we introduce GCDT that is a generalization of $2$d CDT such that baby universes and wormholes are introduced so as to be compatible with the foliation, 
and determine the $1$d effective theory obtained through the sum over all genera. 
In particular, we explain in detail that quantizing $2$d projectable HL gravity with a bi-local interaction yields the $1$d effective theory, and discuss Coleman's mechanism in $2$d CDT. 
Sec.~\ref{sec:summary} is devoted to summary.

\section{$2$d causal dynamical triangulations}
\label{sec:2dcdt}
Two-dimensional causal dynamical triangulations ($2$d CDT) \cite{Ambjorn:1998xu} is a lattice model of quantum geometries based on Regge's discretization \cite{Regge:1961px}. 
In this section, we give an overview of $2$d CDT, and in particular explain how to obtain the quantum Hamiltonian through the continuum limit.    

\begin{figure}[h]
\centering
\includegraphics[scale=.6]{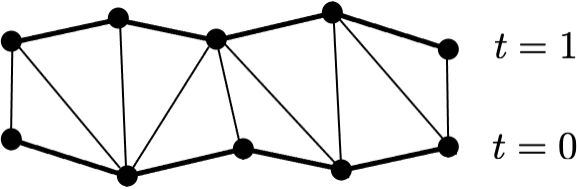}
\caption{A triangulation of a strip: Thick and thin lines are space-like and time-like edges, respectively.}
\label{fig:triangulationofstrip}       
\end{figure}

We start with a two-dimensional globally hyperbolic manifold equipped with a global time foliation: 
\[
\mathcal{M} = \bigcup_{t\in \mathbb{R}} \Sigma_t\ , 
\label{eq:m}
\]
where each leaf $\Sigma_t$ is a one-dimensional Cauchy ``surface'' (line). 
One approximates the manifold with a foliation in such a way that the continuous label $t$ is discretized by integers, i.e. $t\in \mathbb{Z}$;  
each leaf (line) is partitioned by vertices connected by isometric edges; 
vertices among neighboring time steps are connected by isometric edges to form a triangulation of strip (see Fig.~\ref{fig:triangulationofstrip}). 
The edges at a given time step and those connecting vertices in different time steps are, respectively, space-like and time-like edges 
since the squared edge lengths of the space-like edge $a^2_s$ and the time-like edge $a^2_t$ are given by 
\[
a^2_s = \varepsilon^2\ , \ \ \ a^2_t = - \alpha \varepsilon^2\ ,
\label{eq:latticespacing}
\]  
where $\alpha$ is a positive number and $\varepsilon$ is a lattice spacing that serves as a UV cutoff.

$2$d CDT deals with a set of restricted class of Lorentzian triangulations as discussed above. 
In particular, we consider that the topology of the one-dimensional universe (a graph consisting of vertices and edges at a given time) is either $S^1$ or $[0,1]$, 
and the topology will not change during (discrete) time propagation. 
Since the topology is fixed the curvature term plays no role in two dimensions and only the discrete analogue of the cosmological constant term, $\Lambda_0 \int d^2x \sqrt{-g}$ is used as the lattice action of $2$d CDT: 
\[
S_T [\lambda, \alpha] 
= - \frac{\lambda}{\varepsilon^2} 
\left( 
\frac{\sqrt{4\alpha +1}}{4}\varepsilon^2\ n(T)  
\right)\ , 
\label{eq:action}
\]      
where $\lambda/\varepsilon^2$ is the bare cosmological constant with the dimensionless number $\lambda$, 
$n(T)$ the number of triangles in a triangulation $T$,  
and the term inside the parentheses denotes the total area of the triangulation. 
It is useful to rotate to the Euclidean signature which can be performed by changing $\alpha \to - \alpha -i0$. 
Accordingly the lattice action (\ref{eq:action}) changes as follows:
\[
iS_T [\lambda, \alpha] 
\to 
iS_T [\lambda, -\alpha -i0] = - \lambda \frac{\sqrt{4\alpha -1}}{4} n(T)
\equiv - \lambda n(T)
\ , 
\label{eq:rotation}
\]  
where $\alpha$ is chosen to be greater than $1/4$; otherwise, the triangle inequalities will not be satisfied after the rotation.  
In any case we have absorbed the parameter $\alpha$ by the redefinition of the dimensionless cosmological constant $\lambda$. 

The amplitude of the one-dimensional universe that starts with $\ell_1$ edges and ends up with $\ell_2$ edges after the discrete time step $t$ is 
given by the sum over all allowed triangulations: 
\[
G^{(a)}_{\lambda} (\ell_1, \ell_2; t) 
= \sum_{T\in \mathcal{T}^{(a)}(\ell_1,\ell_2,t)} e^{-\lambda n(T)}
= \sum_n e^{-\lambda n} \mathcal{N}^{(a)}(\ell_1,\ell_2,n)
\ , 
\label{eq:amplitude}
\]  
where $\mathcal{T}^{(a)}$ is a set of triangulations whose topology is $[0,1] \times [0,1]$ for $a=0$ and $S^1 \times [0,1]$ for $a=1$, 
and 
\[
\mathcal{N}^{(a)}(\ell_1,\ell_2,n) 
= \# \left\{ T \in \mathcal{T}^{(a)} \ \bigl{|}\ n(T)=n \right\}\ . 
\label{eq:numberoftriangulations}
\]  
When defining the amplitude (\ref{eq:amplitude}), we do not allow the one-dimensional universe to vanish during the discrete time propagation. 
For later convenience, we introduce a marked amplitude: 
\[
G^{(-1)}_{\lambda} (\ell_1, \ell_2; t) 
= \ell_1 G^{(1)}_{\lambda} (\ell_1, \ell_2; t)\ , 
\label{eq:markedamplitude}
\]
where one of the edges in the initial one-dimensional universe is marked. This is because there exist $\ell_1$ possible ways of marking the edges.   
The three kinds of amplitude should satisfy the composition law: 
\[
G^{(1)}_{\lambda} (\ell_1, \ell_2; t_1+t_2) 
&= \sum^{\infty}_{\ell=1} G^{(1)}_{\lambda} (\ell_1, \ell; t_1)\ \ell\ G^{(1)}_{\lambda} (\ell, \ell_2; t_2)\ , \\
G^{(0)}_{\lambda} (\ell_1, \ell_2; t_1+t_2) 
&= \sum^{\infty}_{\ell=1} G^{(0)}_{\lambda} (\ell_1, \ell; t_1)\ G^{(0)}_{\lambda} (\ell, \ell_2; t_2)\ , \\
G^{(-1)}_{\lambda} (\ell_1, \ell_2; t_1+t_2) 
&= \sum^{\infty}_{\ell=1} G^{(-1)}_{\lambda} (\ell_1, \ell; t_1)\  G^{(-1)}_{\lambda} (\ell, \ell_2; t_2)\ , 
\label{eq:compositionlaw}
\]
where for $a=1$ one needs to multiply the amputated amplitudes by $\ell$ since there exist $\ell$ possible ways of gluing to recover the whole amplitude.

It is convenient to introduce the generating function of the number of triangulations. 
Using the notation 
\[
g=e^{-\lambda}\ , 
\label{eq:g}
\]
we define the generating function: 
\[
\widetilde{G}^{(a)} (g,x,y;t) 
&= \sum^{\infty}_{\ell_1=1} \sum^{\infty}_{\ell_2=1} x^{\ell_1} y^{\ell_2} G^{(a)}_{\lambda} (\ell_1, \ell_2; t)\notag \\
&= \sum^{\infty}_{\ell_1=1} \sum^{\infty}_{\ell_2=1} \sum_n x^{\ell_1} y^{\ell_2} g^{n} \mathcal{N}^{(a)}(\ell_1,\ell_2,n)\ ,
\label{eq:generatingfunction} 
\]  
where in the context of quantum gravity, $x$ and $y$ are related to the boundary cosmological constants, $\lambda_1$ and $\lambda_2$, that control the size of the boundaries: 
\[
x=e^{-\lambda_1}\ , \ \ \ y=e^{-\lambda_2}\ . 
\label{eq:xy}
\]
One can reconstruct the amplitude from the generating function through the following relation:
\[
G^{(a)}_{\lambda} (\ell_1, \ell_2; t) 
= \oint_{\mathcal{C}_1} \frac{dx}{2\pi i x^{\ell_1 +1}} \oint_{\mathcal{C}_2} \frac{dy}{2\pi i y^{\ell_2 +1}}\ \widetilde{G}^{(a)} (g,x,y;t)\ , 
\label{eq:amplitudefromgeneratingfunction}
\]
where the contour $\mathcal{C}_1$ ($\mathcal{C}_2$) is chosen to enclose $x=0$ ($y=0$) and to ensure the convergence of $\widetilde{G}^{(a)} (g,x,y;t)$. 
One can derive the relation (\ref{eq:amplitudefromgeneratingfunction}) using the identity:
\[
\oint_{\mathcal{C}} \frac{dz}{2\pi i z^{n +1}} = \delta_{n,0}\ , \ \ \ (n \in \mathbb{Z})\ ,
\]   
where the contour $\mathcal{C}$ encloses $z=0$. 
We provide the composition law for the generating function when $a=0,-1$ in preparation for later calculations:
\[
\widetilde{G}^{(a)} (g,x,y;t_1+t_2) 
= \oint_{\mathcal{C}} \frac{dz}{2\pi i z} 
\widetilde{G}^{(a)} (g,x,z^{-1};t_1)\  
\widetilde{G}^{(a)} (g,z,y;t_2)\ ,\ \  (a=0,-1)\ , 
\label{eq:compositionlaw2}
\]
where the contour encloses $z=0$, and for fixed $g$, $x$ and $y$ lies inside the radius of convergence 
for $\widetilde{G}^{(a)} (g,x,z^{-1};t_1)$ as the series in $1/z$ and for $\widetilde{G}^{(a)} (g,z,y;t_2)$ as the series in $z$, 
which is possible as we will see.   

In what follows, we will discuss the one-step amplitude $G^{(a)}_{\lambda} (\ell_1, \ell_2; 1)$. 
This is because it becomes an important object when computing the whole amplitude.

\subsection{Counting triangulations}
\label{subsec:countingtriangulations}

In this section we focus on the one-step amplitude $G^{(a)}_{\lambda} (\ell_1, \ell_2; 1)$ which is the sum over triangulations of a strip as shown in Fig.~\ref{fig:triangulationofstrip}: 
\[
G^{(a)}_{\lambda} (\ell_1, \ell_2; 1) 
= e^{-\lambda (\ell_1 + \ell_2)} \mathcal{N}^{(a)}(\ell_1,\ell_2,n=\ell_1+\ell_2)\ , 
\label{eqonesteoamplitude} 
\]
and count the number of triangulations $\mathcal{N}^{(a)}(\ell_1,\ell_2,n=\ell_1+\ell_2)$. 
Based on simple combinatorics, one can calculate the case of $a=1$ which has the $S^1 \times [0,1]$ topology:
\[
\mathcal{N}^{(1)}(\ell_1,\ell_2,n=\ell_1+\ell_2) 
= \frac{1}{\ell_1 + \ell_2} 
\begin{pmatrix}
\ell_1 + \ell_2 \\
\ell_1
\end{pmatrix}
= \frac{(\ell_1 + \ell_2-1)!}{\ell_1 ! \ell_2 !}\ .
\label{eq:n1}
\]
Because of the property (\ref{eq:markedamplitude}), 
one can easily compute the case of $a=-1$ that the topology is $S^1 \times [0,1]$ and one of the edges in the initial one-dimensional universe is marked: 
\[
\mathcal{N}^{(-1)}(\ell_1,\ell_2,n=\ell_1+\ell_2) 
= \ell_1 \mathcal{N}^{(1)}(\ell_1,\ell_2,n=\ell_1+\ell_2) 
= \frac{\ell_1 (\ell_1 + \ell_2-1)!}{\ell_1 ! \ell_2 !}\ .
\label{eq:n-1}
\] 
Concerning the case of $a=0$ whose topology is $[0,1] \times [0,1]$, there exist several possibilities depending on the restriction on the leftmost and rightmost triangles. 
If the rightmost triangle is the upward triangle (downward triangle) and the leftmost triangles is the downward triangle (upward triangle), then the counting of triangulations yields 
\[
\mathcal{N}^{(0)}(\ell_1,\ell_2,n=\ell_1+\ell_2) 
= \begin{pmatrix}
\ell_1 + \ell_2 -2 \\
\ell_1 - 1
\end{pmatrix}
= \frac{(\ell_1 + \ell_2-2)!}{(\ell_1 -1) ! (\ell_2 -1) !}\ .
\label{eq:n0}
\] 
In the following, we will use eq.~(\ref{eq:n0}) in the case of $a=0$ for computational simplicity\footnote{
If we do not impose any restriction on the leftmost and rightmost triangles, the number of triangulations becomes 
$\mathcal{N}^{(0)}(\ell_1,\ell_2,n=\ell_1+\ell_2) = ( \ell_1 + \ell_2 )!/(\ell_1! \ell_2!)$. 
}. 

The one-step generating functions can be derived inserting eqs.~(\ref{eq:n1}), (\ref{eq:n-1}) and (\ref{eq:n0}) into eq.~(\ref{eq:generatingfunction}): 
\[
\widetilde{G}^{(1)} (g,x,y;1) 
&= \sum^{\infty}_{\ell_1 = 1} \sum^{\infty}_{\ell_2=1} x^{\ell_1} y^{\ell_2} g^{\ell_1+\ell_2} \mathcal{N}^{(1)}(\ell_1,\ell_2,n=\ell_1+\ell_2) \notag \\
&= - \ln \left( 
\frac{1-gx-gy}{(1-gx)(1-gy)}
\right)\ ; 
\label{eq:generatingfunction1}\\
\widetilde{G}^{(-1)} (g,x,y;1) 
&= \sum^{\infty}_{\ell_1 = 1} \sum^{\infty}_{\ell_2=1} x^{\ell_1} y^{\ell_2} g^{\ell_1+\ell_2} \mathcal{N}^{(-1)}(\ell_1,\ell_2,n=\ell_1+\ell_2) \notag \\
&= \frac{g^2xy}{(1-gx)(1-gx-gy)} \ ; 
\label{eq:generatingfunction-1}\\
\widetilde{G}^{(0)} (g,x,y;1) 
&= \sum^{\infty}_{\ell_1 = 1} \sum^{\infty}_{\ell_2=1} x^{\ell_1} y^{\ell_2} g^{\ell_1+\ell_2} \mathcal{N}^{(0)}(\ell_1,\ell_2,n=\ell_1+\ell_2) \notag \\
&= \frac{g^2xy}{1-gx-gy} \ .
\label{eq:generatingfunction0}
\]
In fact, one can also obtain eq.~(\ref{eq:generatingfunction-1}) through $\widetilde{G}^{(-1)} (g,x,y;1) = x \frac{\partial}{\partial x} \widetilde{G}^{(1)} (g,x,y;1)$. 

Alternatively, it is possible to compute the one-step generating functions directly by simple combinatorics: 
\[
\widetilde{G}^{(1)} (g,x,y;1) 
&= \sum^{\infty}_{s=1} \frac{1}{s} \left( \sum^{\infty}_{k=1} (gx)^k \sum^{\infty}_{l=1} (gy)^l  \right)^s 
= - \ln \left( 
\frac{1-gx-gy}{(1-gx)(1-gy)}
\right)\ ; 
\label{eq:generatingfunction1v2}\\
\widetilde{G}^{(-1)} (g,x,y;1) 
&= \sum^{\infty}_{k=0} \left( gx \sum^{\infty}_{l=0} (gy)^l \right)^k - \sum^{\infty}_{k=0} (gx)^k 
= \frac{g^2xy}{(1-gx)(1-gx-gy)} \ ; 
\label{eq:generatingfunction-1v2}\\
\widetilde{G}^{(0)} (g,x,y;1) 
&= \sum^{\infty}_{s=1}  \left( \sum^{\infty}_{k=1} (gx)^k \sum^{\infty}_{l=1} (gy)^l  \right)^s 
= \frac{g^2xy}{1-gx-gy} \ .
\label{eq:generatingfunction0v2}
\]

\subsection{Continuum limit}
\label{subsec:continuumlimit}
   
All is now set for computing the amplitude in the continuum limit. 
In this section, however, instead of directly computing the amplitude in the continuum limit, 
we will derive the differential equation that the continuum amplitude satisfies. 

Before going into details any further, let us explain some basic facts of the continuum limit. 
In order to remove the cutoff $\varepsilon$ through the continuum limit, one has to tune the bare coupling constants $(g,x,y)$ to their critical values $(g_c,x_c,y_c)$. 
At the critical values, the generating function hits the radii of convergence and therefore becomes non-analytic. 
Approaching such a critical point, infinitely many triangles and boundary edges become important in the summation of the generating function, 
i.e., essentially, the average number of triangles and boundary edges become infinity at the critical point. 
Having this in mind, one may intuitively understand that the continuous surface would be obtained if $(g,x,y) \to (g_c,x_c,y_c)$ and $\varepsilon \to 0$ in a correlated manner. 

Introducing $\lambda_c = - \ln [g_c]$, $\lambda_{1c} = - \ln [x_c]$ and $\lambda_{2c} = - \ln [y_c]$, 
one can transmute the dimension of the lattice spacing $\varepsilon$ into the dimension of the renormalized bulk and boundary cosmological constants through the continuum limit:  
\[
\Lambda = \lim_{\substack{\lambda \to \lambda_c \\ \varepsilon \to 0}} \frac{\lambda - \lambda_c}{\varepsilon^2}\ , \ \ \ 
X = \lim_{\substack{\lambda_1 \to \lambda_{1c} \\ \varepsilon \to 0}} \frac{\lambda_1 - \lambda_{1c}}{\varepsilon}\ , \ \ \ 
Y = \lim_{\substack{\lambda_2 \to \lambda_{2c} \\ \varepsilon \to 0}} \frac{\lambda_2 - \lambda_{2c}}{\varepsilon}\ ,
\label{eq:renormalizedcc}
\]         
where $\Lambda$ is the renormalized bulk cosmological constant, and $X$ and $Y$ are the renormalized boundary cosmological constants. 
Therefore, the divergent bare cosmological constants get additive renormalizations so as to obtain the finite renormalized cosmological constants that set the scale at IR.   

In the following, we discuss the continuum limit in detail with respect to each topology of spacetime. 

\subsubsection{$S^1 \times [0,1]$ topology}
\label{subsubsec:s1times01}

We consider the case of $a=-1$, i.e. $S^1 \times [0,1]$ topology with a marked boundary.   
Using the composition law (\ref{eq:compositionlaw2}) and the one-step generating function (\ref{eq:generatingfunction-1v2}), 
one obtains
\[
\widetilde{G}^{(-1)} (g,x,y;t+1) 
&= \oint_{\mathcal{C}} \frac{dz}{2\pi i z} 
\widetilde{G}^{(-1)} (g,x,z^{-1};1)\  
\widetilde{G}^{(-1)} (g,z,y;t)\notag \\
&=\oint_{\mathcal{C}} \frac{dz}{2\pi i} 
\frac{g^2x}{(1-gx)^2 (z-g/(1-gx))} \frac{\widetilde{G}^{(-1)} (g,z,y;t)}{z} \notag\\
&= \frac{gx}{1-gx}\ \widetilde{G}^{(-1)} \left( g, \frac{g}{1-gx} ,y;t \right)\ . 
\label{eq:-1relation}
\]
In the last equality, we have picked up a pole at $z=g/(1-gx)$, and there exists no pole at $z=0$ since $\frac{\widetilde{G}^{(-1)} (g,z,y;t)}{z}$ is regular. 
Through iterative use of eq.~(\ref{eq:-1relation}), one can analytically compute the generating function, and extract the information of the critical point \cite{Ambjorn:1998xu}.       
However, we do not compute the generating function directly to obtain the critical point. 
Instead, we follow the procedure shown in \cite{Ambjorn:2022btk}: 
One assumes the existence of the critical point, and determines the value of the critical coupling constants from the consistency.  

We assume the critical point characterized by the critical coupling constants $(g_c,x_c,y_c)$ 
and use the following parametrization:
\[
g=g_c e^{-\varepsilon^2 \Lambda}, \ \ \ 
x=x_c e^{-\varepsilon X}\ , \ \ \ 
y=y_c e^{-\varepsilon Y}\ . 
\label{eq:parametrization}
\]
Assuming the scalings 
\[
T = \varepsilon t\ , \ \ \ L_1 = \varepsilon \ell_1\ , \ \ \ L_2 = \varepsilon \ell_2\ , 
\label{eq:continuumtandl}
\]
we introduce the renormalized amplitude and the renormalized generating function at the critical point by the multiplicative renormalizations:
\[
G^{(-1)}_{\Lambda} (L_1,L_2; T) 
&= \lim_{\varepsilon \to 0} C_{\varepsilon}\ G^{(-1)}_{\lambda} (\ell_1,\ell_2; t)\ , \label{eq:continuumg}\\
\widetilde{G}^{(-1)}_{\Lambda} (X,Y; T) 
&= \lim_{\varepsilon \to 0} \widetilde{C}_{\varepsilon}\ \widetilde{G}^{(-1)} (g, x, y; t)\ , \label{eq:continuumgtilde}
\]
where $C_{\varepsilon}$ and $\widetilde{C}_{\varepsilon}$ are real functions of $\varepsilon$ that will be fixed later.   
One can determine the critical coupling constants for the consistency. 
Using the scaling behavior (\ref{eq:continuumgtilde}), eq.~(\ref{eq:-1relation}) can yield the sensible continuum limit if the critical coupling constants satisfy 
\[
\frac{g_c x_c}{1-g_c x_c} = 1\ , \ \ \ \frac{g_c}{1-g_c x_c} = x_c \ \ \ 
\Rightarrow \ \ \ 
g_c = \frac{1}{2}\ , \ \ \ x_c = 1\ .
\]  
Let us determine the function $C_{\varepsilon}$ in such a way that 
the composition law (\ref{eq:compositionlaw}) holds in the continuum limit as 
\[
G^{(-1)}_{\Lambda} (L_1,L_2; T_1 + T_2) 
= \int^{\infty}_{0}dL\  
G^{(-1)}_{\Lambda} (L_1,L; T_1) G^{(-1)}_{\Lambda} (L,L_2; T_2)\ , 
\label{eq:compositionlawcontinuum} 
\]
which is possible if $C_{\varepsilon} = \varepsilon^{-1}$. 
The function $\widetilde{C}_{\varepsilon}$ can be determined in such way that eq.~(\ref{eq:generatingfunction}) makes sense in the continuum limit, i.e. 
 \[
\widetilde{G}^{(-1)}_{\Lambda} (X,Y; T) 
= \int^{\infty}_{0} dL_1 \int^{\infty}_{0}dL_2\ e^{-XL_1} e^{-YL_2}  
G^{(-1)}_{\Lambda} (L_1,L_2; T)\ , 
\label{eq:continuumlaplacetr}
\]
which is possible if $\widetilde{C}_{\varepsilon} = \varepsilon$.

Now we wish to take the continuum limit of eq.~(\ref{eq:-1relation}). 
For notational convenience, we redefine the renormalized coupling constants as follows:
\[
g= \frac{1}{2} e^{-\varepsilon^2 \Lambda} \equiv \frac{1}{2} \left( 1 - \frac{1}{2}\varepsilon^2 \Lambda \right), \ \ \ 
x= e^{-\varepsilon X} \equiv 1 - \varepsilon X \ , \ \ \ 
y= e^{-\varepsilon Y} \equiv 1 - \varepsilon Y \ . 
\label{eq:parametrization2}
\]
Plugging eqs.~(\ref{eq:continuumtandl}), (\ref{eq:parametrization2}) into eq.~(\ref{eq:-1relation}), 
one obtains the differential equation: 
\[
\frac{\partial}{\partial T} \widetilde{G}^{(-1)}_{\Lambda} (X,Y; T)
= - \frac{\partial}{\partial X} \left[ (X^2-\Lambda) \widetilde{G}^{(-1)}_{\Lambda} (X,Y; T) \right]\ . 
\label{eq:differentialeq-1}
\]
Doing a little math, one can also derive the continuum description of eq.~(\ref{eq:amplitudefromgeneratingfunction}):
\[
G^{(-1)}_{\Lambda} (L_1,L_2; T) 
= \int^{c+i\infty}_{c-i\infty} \frac{dX}{2\pi i} \int^{c+i\infty}_{c-i\infty} \frac{dY}{2\pi i}\
e^{L_1 X} e^{L_2 Y}\ \widetilde{G}^{(-1)}_{\Lambda} (X,Y; T)\ ,
\label{eq:inverselaplacetr} 
\]
where $c$ is a suitable real number. 
Using the inverse Laplace transform (\ref{eq:inverselaplacetr}) and eq.~(\ref{eq:differentialeq-1}), 
one obtains the differential equation that the continuum amplitude satisfies:
\[
\frac{\partial}{\partial T} G^{(-1)}_{\Lambda} (L_1,L_2; T) 
= - \hat{H}^{(-1)} (L_1)\ G^{(-1)}_{\Lambda} (L_1,L_2; T)\ ,  
\label{eq:differentialeq-1v2}
\]
where 
\[
\hat{H}^{(-1)}(L) = - L \frac{\partial^2}{\partial L^2} + \Lambda L\ . 
\label{hamiltonian-1}
\] 
As a result, one can interpret the continuum limit of $2$d CDT as a quantum system of the one-dimensional universe with length $L$ 
that propagates in time $T$ following the quantum Hamiltonian (\ref{hamiltonian-1}). 
The quantum Hamiltonian (\ref{hamiltonian-1}) is Hermitian with respect to the inner product:
\[ 
%\left\langle \phi  | \psi \right\rangle = 
\int^{\infty}_{0} \frac{dL}{L} \phi^* (L)\ (\hat{H}^{(-1)} \psi) (L)
=
\int^{\infty}_{0} \frac{dL}{L} (\hat{H}^{(-1)} \phi)^* (L)\ \psi (L)\ . 
\label{eq:innerproduct-1} 
\]  
 
The differential equation for the un-marked amplitude can be easily read off inserting the continuum limit of eq.~(\ref{eq:markedamplitude})
\[
G^{(-1)}_{\Lambda} (L_1,L_2; T) 
= L_1 G^{(1)}_{\Lambda} (L_1,L_2; T)\ , 
\label{eq:continuummarkedamplitude} 
\]
into eq.~(\ref{eq:differentialeq-1v2}): 
\[
\frac{\partial}{\partial T} G^{(1)}_{\Lambda} (L_1,L_2; T) 
= - \hat{H}^{(1)} (L_1)\ G^{(1)}_{\Lambda} (L_1,L_2; T)\ ,  
\label{eq:differentialeq1v2}
\]
where 
\[
\hat{H}^{(1)}(L) = - \frac{\partial^2}{\partial L^2} L + \Lambda L\ . 
\label{hamiltonian1}
\]   
The quantum Hamiltonian (\ref{hamiltonian1}) is Hermitian with respect to the inner product:
\[
\int^{\infty}_{0} LdL \phi^* (L)\  (\hat{H}^{(1)} \psi) (L) 
=
\int^{\infty}_{0} LdL (\hat{H}^{(1)} \phi)^* (L)\ \psi (L)
\ . 
\label{eq:innerproduct1} 
\]

\subsubsection{$[0,1] \times [0,1]$ topology}
\label{subsubsec:01times01}

Let us consider the case of $a=0$, i.e. $[0,1] \times [0,1]$ topology. 
We basically follow the procedure shown in Sect.~\ref{subsubsec:s1times01}.   
Using the composition law (\ref{eq:compositionlaw2}) and the one-step generating function (\ref{eq:generatingfunction0v2}), 
one obtains
\[
\widetilde{G}^{(0)} (g,x,y;t+1) 
&= \oint_{\mathcal{C}} \frac{dz}{2\pi i z} 
\widetilde{G}^{(0)} (g,x,z^{-1};1)\  
\widetilde{G}^{(0)} (g,z,y;t)\notag \\
&=\oint_{\mathcal{C}} \frac{dz}{2\pi i} 
\frac{g^2x}{(1-gx)(z-g/(1-gx))} 
\frac{\widetilde{G}^{(0)} (g,z,y;t)}{z} \notag\\
&= gx\ \widetilde{G}^{(0)} \left( g, \frac{g}{1-gx} ,y;t \right)\ . 
\label{eq:0relation}
\]
From eq.~(\ref{eq:0relation}), one may obtain a sensible continuum limit 
if the critical coupling constants are the same as before, i.e. $(g_c,x_c,y_c)=(1/2,1,1)$, and if the multiplicative renormalization is treated carefully:  
\[
\widetilde{G}^{(0)}_{\Lambda} (X,Y; T) 
= \lim_{\varepsilon \to 0}  \frac{\varepsilon}{2^t}  \widetilde{G}^{(0)} (g, x, y; t)\ . 
\label{eq:continuumgtilde0}
\]
In fact, this assumption yields the correct continuum limit. 
Plugging eqs.~(\ref{eq:continuumtandl}), (\ref{eq:parametrization2}) into eq.~(\ref{eq:0relation}) and using eq.~(\ref{eq:continuumgtilde0}), 
one obtains the differential equation:
\[
\frac{\partial}{\partial T} \widetilde{G}^{(0)}_{\Lambda} (X,Y; T)
= - \left( X + (X^2 - \Lambda) \frac{\partial}{\partial X} \right) \widetilde{G}^{(0)}_{\Lambda} (X,Y; T) \ . 
\label{eq:differentialeq0}
\] 
Defining the continuum amplitude in such a way that the inverse Laplace transform (\ref{eq:inverselaplacetr}) holds in the case of $a=0$ as well, i.e. 
\[
G^{(0)}_{\Lambda} (L_1,L_2; T) 
= \int^{c+i\infty}_{c-i\infty} \frac{dX}{2\pi i} \int^{c+i\infty}_{c-i\infty} \frac{dY}{2\pi i}\
e^{L_1 X} e^{L_2 Y}\ \widetilde{G}^{(0)}_{\Lambda} (X,Y; T)\ ,
\label{eq:inverselaplacetr0} 
\]
and using eq.~(\ref{eq:inverselaplacetr0}), the differential equation (\ref{eq:differentialeq0}) becomes 
\[
\frac{\partial}{\partial T} G^{(0)}_{\Lambda} (L_1,L_2; T) 
= - \hat{H}^{(0)} (L_1)\ G^{(0)}_{\Lambda} (L_1,L_2; T)\ ,  
\label{eq:differentialeq0v2}
\]
where $\hat{H}^{(0)}$ is the quantum Hamiltonian obtained in Refs.~\cite{DiFrancesco:2000nn, Durhuus:2001sp}:  
\[
\hat{H}^{(0)}(L) = - \frac{\partial}{\partial L} L \frac{\partial}{\partial L} + \Lambda L\ . 
\label{hamiltonian0}
\]   
The quantum Hamiltonian (\ref{hamiltonian0}) is hermitian with respect to the inner product:
\[
%\left\langle \phi  | \psi \right\rangle = 
\int^{\infty}_{0} dL \phi^* (L)\ (\hat{H}^{(0)}\psi) (L)
=
\int^{\infty}_{0} dL (\hat{H}^{(0)} \phi)^* (L)\ \psi (L)
\ . 
\label{eq:innerproduct0} 
\]

\subsection{Short summary of $2$d CDT}
\label{subsec:shortsummaryof2dcdt}
As discussed in Sect.~\ref{subsec:continuumlimit}, the continuum limit of $2$d CDT is described by the quantum mechanics of a one-dimensional universe with length $L$ that propagates in time $T$ 
based on the Hamiltonian $\hat{H}^{(a)}$: 
\[
\hat{H}^{(-1)} = - L \frac{\partial^2}{\partial L^2} + \Lambda L\ , \ \ \ 
\hat{H}^{(1)} = - \frac{\partial^2}{\partial L^2} L + \Lambda L\ , \ \ \
\hat{H}^{(0)} = - \frac{\partial}{\partial L} L \frac{\partial}{\partial L} + \Lambda L\ ,
\label{eq:hamiltonians}
\]   
where the label $a$ classifies the topology of the one-dimensional universe: 
$S^1$ and $[0,1]$ for $a=1$ and $a=0$, respectively. 
When $a=-1$, the closed one-dimensional universe is marked. 
Let us define the eigenstates of $L$ as $|L\rangle_{a}$ that satisfy 
the completeness relation:  
\[
\hat{1} = \int^{\infty}_{0}L^{a}dL\ |L\rangle_{a} {}_{a}\langle L | \ \ \ 
\Leftrightarrow 
\ \ \ 
{}_a \langle L' | L \rangle_{a} = \frac{1}{L^a} \delta (L - L')\ . 
\label{eq:completenessrelations}
\] 
Note that $|L \rangle_{-1} = L | L \rangle_1$. 
One can then express the amplitudes as matrix elements:
\[
G^{(1)}_{\Lambda} (L_1, L_2; T) 
&= {}_{1}\langle L_2 | e^{-T \hat{H}^{(1)}}  |L_1 \rangle_{1}\ , \\
G^{(-1)}_{\Lambda} (L_1, L_2; T) 
&= {}_{1}\langle L_2 | e^{-T \hat{H}^{(-1)}}  |L_1 \rangle_{-1}\ , \\
G^{(0)}_{\Lambda} (L_1, L_2; T) 
&= {}_{0}\langle L_2 | e^{-T \hat{H}^{(0)}}  |L_1 \rangle_{0}\ , 
\label{eq:matrixelement2}
\] 
Using eq.~(\ref{eq:completenessrelations}), one can show that the composition laws hold: 
For $a=-1,0$, 
\[
G^{(a)}_{\Lambda} (L_1,L_2; T_1 + T_2) 
= \int^{\infty}_{0}dL\  
G^{(a)}_{\Lambda} (L_1,L; T_1) G^{(a)}_{\Lambda} (L,L_2; T_2)\ ,
\label{eq:compositionlawcontinuum_a-10} 
\]
and for $a=1$,  
\[
G^{(1)}_{\Lambda} (L_1,L_2; T_1 + T_2) 
&= \int^{\infty}_{0}dL\  
G^{(1)}_{\Lambda} (L_1,L; T_1) L G^{(1)}_{\Lambda} (L,L_2; T_2)\ . 
\label{eq:compositionlawcontinuum_a1}
\]

\section{$2$d projectable Ho\v rava-Lifshitz quantum gravity}
\label{sec:2dHLQG}

We wish to introduce the classical field theory that reproduces the continuum limit of $2$d CDT once it is quantized. 
The field theory is a certain version of the two-dimensional Ho\v rava-Lifshitz gravity ($2$d HL gravity).

The starting point is the same class of manifold with a foliation (\ref{eq:m}) where $\Sigma_t$ is a one-dimensional space labelled by $t$: 
\[
\Sigma_t = \{ x^{\mu} \in \mathcal{M}\ |\ f(x^{\mu}) = t \}\ , \ \ \ \text{with} \ \ \ \mu = 0,1\ . 
\label{eq:sigma} 
\] 
Choosing that $f(x^{\mu}) =x^0$, the time direction can be decomposed into the two directions, i.e. the normal and the tangential to $\Sigma_t$: 
\[
\left( \partial_t \right)^{\mu} 
= \frac{\partial x^{\mu}}{\partial t} 
= N n^{\mu} + N^1 E^{\mu}_{1}\ , 
\label{eq:timedirection}
\] 
where $n^{\mu}$ and $E^{\mu}_1$ are, respectively, the unit normal vector and the tangent vector defined as 
\[
n^{\mu} = 
\left(
\frac{1}{N}, - \frac{N^1}{N}
\right)\ , \ \ \ 
E^{\mu}_1 = \delta^{\mu}_1\ . 
\label{eq:normalandtangent}
\]  
Here $N$ and $N^1$ are the Lapse function and the shift vector. 
Through the use of eq.~(\ref{eq:timedirection}), 
one can parametrize the metric $g_{\mu \nu}$ on $\mathcal{M}$ as follows:
\[
ds^2 
= g_{\mu \nu}  dx^{\mu} dx^{\nu} 
= - N^2  dt^2 + h_{11}  \left(dx +N^1 dt \right) \left( dx +N^1 dt \right) \ , 
\label{eq:admmetric}
\]  
where $t=x^0$ and $x=x^1$; $h_{11}$ is the spatial metric on $\Sigma_t$ defined as $h_{11} = E^{\mu}_{1} E^{\nu}_1 g_{\mu \nu}$. 

$2$d HL gravity is a field theory that preserves the structure of the time foliation, or in other words, it is invariant under 
the foliation-preserving diffeomorphisms (FPD):      
\[
t \to t + \xi^0 (t)\ , \ \ \ x \to x + \xi^1 (t,x)\ . 
\label{eq:fpd}
\]
The fields transform under FPD as follows:
\[
\delta_{\xi} h_{11} &= \xi^0 \partial_0 h_{11} + \xi^1 \partial_1 h_{11} + 2h_{11}\partial_1 \xi^1\ , \label{eq:deltah11}\\
\delta_{\xi} N_1 &= \xi^{\mu} \partial_{\mu} N_1 + N_1 \partial_{\mu} \xi^{\mu} + h_{11} \partial_0 \xi^1\ , \label{eq:deltan1}\\
\delta_{\xi} N &= \xi^{\mu}\partial_{\mu} N + N \partial_0 \xi^0\ . \label{eq:deltan}
\] 
where $N_1 = h_{11}N^1$. 
Here if a function is a constant on each foliation $\Sigma_t$, such a function is called projectable. 
In fact, implementing FPD the projectable Lapse function, i.e. $N=N(t)$, stays as a function only of time.  
The HL gravity with the projectable Lapse function is dubbed the projectable HL gravity. 
Since it is $2$d projectable HL gravity that reproduces the continuum limit of $2$d CDT once it is quantized, 
from now we focus on this special version of HL gravity.   

The action of $2$d projectable HL gravity is given by 
\[
I =
\int dt\ \mathcal{L}
=\frac{1}{\kappa} 
\int dtdx\ 
N(t) \sqrt{h(t,x)} 
\left(
(1-\eta) K^2(t,x) - 2 \widetilde{\Lambda}
\right)\ , 
\label{eq:hlaction}
\]
where $\mathcal{L}$ is the Lagrangian; 
$\eta$, $\widetilde{\Lambda}$ and $\kappa$ are a dimensionless parameter, 
the cosmological constant and the (dimensionless) gravitational coupling constant, respectively; 
$h$ is the determinant of the spatial metric $h_{11}$, i.e. $h=h_{11}$; 
$K$ is the trace of the extrinsic curvature $K_{11}$ given by 
\[
K_{11} = \frac{1}{2N} \left( \partial_0 - 2\nabla_1 N_1 \right)\ , \ \ \ 
\text{with}\ \ \ 
\nabla_1 N_1 = \partial_1 N_1 - \Gamma^1_{11} N_1\ .
\label{eq:extrinsiccurvature}
\] 
Here $\Gamma^1_{11}$ is the spatial Christoffel symbol: 
\[
\Gamma^1_{11} 
= \frac{1}{2} h^{11} \partial_1 h_{11}\ . 
\label{eq:christoffel}
\]
One can in principle add higher spatial derivative terms to the action (\ref{eq:hlaction}). 
However, such terms are not necessary because the model is renormalizable in two dimensions without introducing them, 
and therefore we omit such terms. 

The continuum limit of $2$d CDT can be precisely obtained if quantizing $2$d projectable HL gravity 
with the following identification of parameters: 
\[
\Lambda = \frac{ \widetilde{\Lambda} }{2(1-\eta)}\ , \ \ \ 
\eta < 1\ , \ \ \ 
\kappa = 4(1-\eta)\ , 
\label{eq:parameteridentification}
\]
where $\Lambda$ is the renormalized cosmological constant of CDT defined by eq.~(\ref{eq:parametrization2}).

\subsection{Quantization}
\label{sec:quantization}

Let us overview the quantization of $2$d projectable HL gravity shown in Ref.~\cite{Ambjorn:2013joa} (see also Ref.~\cite{Li:2014bla} for another article examining this issue). 

We introduce the conjugate momentum of $\sqrt{h}$ as $\Pi$, which satisfy the Poisson bracket: 
\[
\left\{ 
\sqrt{h(t,x)}, \Pi (t,x')
\right\}
= \delta (x-x')\ . 
\label{eq:poisson}
\] 
Through the Legendre transformation of the Lagrangian (\ref{eq:hlaction}), 
one obtains the Hamiltonian of $2$d projectable HL gravity:
\[
H = \int dx
\left( 
\Pi(t,x) \partial_t \sqrt{h (t,x)} 
\right)
- \mathcal{L} 
= 
N(t) \mathcal{C}(t) +
\int dx\ 
N_1 (t,x) \mathcal{C}^1 (t,x)
\ . 
\label{eq:classicalhamiltonian}
\]
Since $2$d projectable HL gravity is a singular system due to the invariance under FPD, 
there exist two kinds of constraint:  
\[
\mathcal{C}^1 (t,x)  
&= - \frac{\partial_1 \Pi (t,x)}{\sqrt{h (t,x)}} \approx 0\ , \label{eq:momentumconst}\\
\mathcal{C} (t)
&= \int dx\ 
\left(
\frac{\kappa}{4(1-\eta)} \Pi^2(t,x) \sqrt{h(t,x)} + \frac{2}{\kappa} \widetilde{\Lambda} \sqrt{h(t,x)}
\right)
\approx 0 
\ , \label{eq:classicalhamiltonianconst} 
\]  
where $\mathcal{C}^1 (t,x) \approx 0$ is the momentum constraint, 
and $\mathcal{C} (t) \approx 0$ is the Hamiltonian constraint which is global because of the projectable Lapse function\footnote{
The Hamiltonian and the momentum constraints come from the consistency conditions that 
the primary constraints, $\Pi_N \approx 0$ and $\Pi_{N_1} \approx 0$, should be preserved under the time flow 
where $\Pi_N$ and $\Pi_{N_1}$ are the conjugate momenta of $N$ and $N_1$, respectively.   
}.   

The strategy is to solve the momentum constraint (\ref{eq:momentumconst}) at the level of classical theory, i.e. 
\[
\mathcal{C}^1 (t,x) = 0 \ \ \ 
\Rightarrow \ \ \ 
\Pi (t,x) = \Pi (t)\ , 
\label{eq:solvemomentumconst} 
\] 
meaning that the conjugate momentum becomes a function only of time. 
Applying eq.~(\ref{eq:solvemomentumconst}), 
the Hamiltonian (\ref{eq:classicalhamiltonian}) reduces to the one for the one-dimensional system: 
\[
H =
N(t) \left(
\frac{\kappa}{4 (1-\eta)} \Pi^2(t) L(t) 
+ \frac{2}{\kappa} \widetilde{\Lambda} L(t) 
\right)\ ,   
\ \ \
\text{with} \ \ \ 
L(t) = \int dx \sqrt{h(t,x)}\ , 
\label{eq:reducedhamiltonian}
\]  
where $L(t)$ is the invariant length of the one-dimensional universe. 
Let us discuss solutions to the Hamiltonian constraint. 
If $(\eta -1)\widetilde{\Lambda}>0$, one has a solution:
\[
\Pi^2 = \frac{8(\eta - 1)}{\kappa^2}\ \widetilde{\Lambda}\ , 
\label{eq:hamiltonianconstsol1}
\] 
which means that the extrinsic curvature is a constant. 
On the other hand, if $(\eta -1)\widetilde{\Lambda}<0$, 
the only solution is 
\[
L=0\ . 
\label{eq:hamiltonianconstsol2}
\]

Hereafter we apply the parametrization (\ref{eq:parameteridentification}): We choose 
$(\eta -1)\widetilde{\Lambda}<0$, set the unimportant dimensionless gravitational constant as $\kappa = 4(1-\eta)$, and redefine the cosmological constant as $\Lambda = \frac{ \widetilde{\Lambda} }{2(1-\eta)}$.  
Since $\kappa > 0$, this means that $\eta <1$ which selects the correct sign of the kinetic term, and the positive cosmological constant $\widetilde{\Lambda}>0$.  
The dynamics of the classical $1$d system with the Hamiltonian (\ref{eq:reducedhamiltonian}) can be alternatively described by the following action: 
\[
S= \int^{1}_{0} dt \left(
\frac{\dot{L}^2(t)}{4N(t)L(t)} - \Lambda N(t)L(t)
\right)\ , 
\label{eq:reducedaction}
\]  
where $\dot{L}(t) := \frac{d}{dt} L(t)$. 
We then introduce the proper time:
\[
\tau (s) = \int^{s}_{0}dt\ N(t)\ , \ \ \ s\in [0,1]\ .
\label{eq:propertime}
\]
Since the proper time (\ref{eq:propertime}) is invariant under the reparametrization of time, $t\to t + \xi^0 (t)$, 
if one fixes the Lapse function as $N(\tau)=1$, the length of the one-dimensional universe $L(\tau)$ is also invariant under the time redefinition. 
Therefore, it makes sense to discuss the amplitude such that the one-dimensional universe with the length $L_1:=L(\tau=0)$ propagates in the proper time $\tau$, 
and ends up with the universe whose length is given by $L_2 := L_2(\tau = T)$.

With this understanding, we consider such an amplitude based on the path integral. 
For convenience, we rotate $t \to it$, which is possible thanks to the foliation and introduce the Euclidean action:
\[
S_E = 
\int^{1}_{0} dt \left(
\frac{\dot{L}^2(t)}{4N(t)L(t)} + \Lambda N(t)L(t)
\right)\ .  
\label{eq:euclideanaction}
\] 
Using the Euclidean action (\ref{eq:euclideanaction}), the amplitude becomes 
\[
\mathcal{G}_{\Lambda} (L_1,L_2;T)
= \int \frac{\mathcal{D}N(t)}{\text{Diff [0,1]}} \int^{L(1)=L_2}_{L(0)=L_1} \mathcal{D}L(t)\ e^{-S_E [N(t), L(t)]}\ , 
\label{eq:pathintegral}
\]
where 
\[
T := \int^{1}_{0} dt\ N(t)\ . 
\label{eq:T}
\]   
We fix the Lapse function as $N(\tau)=1$ introducing the corresponding Faddeev-Popov (FP) determinant. 
Since the FP determinant only gives an overall constant, we will omit it in the following. 
After the gauge fixing, the amplitude (\ref{eq:pathintegral}) becomes
\[
\mathcal{G}_{\Lambda} (L_1,L_2;T) 
= \int^{L(T)=L_2}_{L(0)=L_1} 
 \mathcal{D}L(\tau)\ 
 \exp \left[ 
 - \int^T_0 d\tau 
 \left(
 \frac{\dot{L}^2(\tau)}{4L(\tau)} + \Lambda L(\tau)
 \right)
 \right]\ , 
 \label{eq:pathintegral2}
\]     
where $\dot{L}(\tau) := \frac{d}{d \tau} L(\tau)$. 

So far, we have not specified the integral measure $\mathcal{D}L(\tau)$. 
We apply the three kinds of measure given by 
\[
\mathcal{D}^{(a)}L(\tau) 
= \prod^{\tau=T}_{\tau=0} L^a(\tau)\ dL(\tau)\ , \ \ \ (a=0,\pm 1)\ .
\label{eq:measure}
\] 
Accordingly, we consider the three kinds of amplitude, i.e. $\mathcal{G}^{(a)}_{\Lambda}(L_1,L_2;T)$, 
and rewrite them introducing the quantum Hamiltonian $\hat{H}^{(a)}$:
\[
\mathcal{G}^{(a)}_{\Lambda}(L_1,L_2;T) 
= {}_{a}\langle L_2 | e^{-T \hat{H}^{(a)}}  |L_1 \rangle_{a}\ ,
\label{eq:amplitudematrix}
\] 
where the eigenstates of $L$ satisfy the completeness relation: 
\[
1 = \int^{\infty}_{0}L^{a}dL\ |L\rangle_{a} {}_{a}\langle L | \ \ \ 
\Leftrightarrow 
\ \ \ 
{}_a \langle L' | L \rangle_{a} = \frac{1}{L^a} \delta (L - L')\ . 
\label{eq:completenessrelations2}
\] 
In order to read off the quantum Hamiltonian $\hat{H}^{(a)}$, we discretize the proper time interval in steps of $\varepsilon$, 
and calculate the one-step matrix element $\mathcal{G}^{(a)}_{\Lambda}(L,L';\varepsilon)$. 
The normalization can be fixed so as to satisfy the following equation:
\[
\lim_{\varepsilon \to 0} \int^{\infty}_{0} L^a dL\ \mathcal{G}^{(a)}_{\Lambda}(L,L';\varepsilon) = 1\ , 
\label{eq:normalization} 
\] 
which comes from the completeness relation (\ref{eq:completenessrelations2}). 
The result is 
\[
\mathcal{G}^{(a)}_{\Lambda}(L,L';\varepsilon)  
= \frac{(LL')^{(1-a)/2}}{L' \sqrt{4\pi \varepsilon L'}} e^{-\frac{(L-L')^2}{4\varepsilon L'} - \Lambda \varepsilon L'}\ . 
\label{eq:onesteppropagator}
\] 
Integrating the one-step amplitude together with a function, $\psi_a (L) = {}_a \langle L | \psi \rangle$, for $\varepsilon \ll 1$, 
one can read off the quantum Hamiltonian:  
\[
\psi_a (L'; \varepsilon) 
&= {}_{a}\langle L' | e^{- \varepsilon \hat{H}^{(a)}}  | \psi \rangle \notag \\ 
&= \int^{\infty}_{0} L^a dL\ {}_{a}\langle L' | e^{- \varepsilon \hat{H}^{(a)}}  |L \rangle_{a} {}_a \langle L | \psi \rangle \notag \\  
& \cong \psi_a (L') - \varepsilon \hat{H}^{(a)}\psi_a (L') + \mathcal{O} (\varepsilon^{3/2})\ .  
\label{eq:readoffhamiltonian} 
\]   
Using eq.~(\ref{eq:onesteppropagator}), one obtains    
\[
\hat{H}^{(-1)} (L) = - L \frac{d^2}{dL^2} + \Lambda L\ , \ \ \ 
\hat{H}^{(0)} (L) =  - \frac{d}{dL} L \frac{d}{dL}  + \Lambda L\ , \ \ \ 
\hat{H}^{(1)} (L) =  - \frac{d^2}{dL^2} L + \Lambda L\ , 
\label{eq:hamiltonianhorava}
\]

The quantum Hamiltonians (\ref{eq:hamiltonianhorava}) obtained by quantizing $2$d projectable HL gravity are 
precisely equivalent to those obtained by the continuum limit of $2$d CDT (see eqs.~(\ref{hamiltonian-1}), (\ref{hamiltonian1}) and (\ref{hamiltonian0})). 
The amplitudes are related as follows:
\[
\mathcal{G}^{(-1)}_{\Lambda}(L,L';T) = L' G^{(-1)}_{\Lambda} (L,L';T)\ , \ \ \  
\mathcal{G}^{(a)}_{\Lambda}(L,L';T) = G^{(a)}_{\Lambda} (L,L';T)\ , \ \ \  (a=0,1)\ . 
\label{eq:relationamplitudes}
\]
Thereby, we understand that the classical field theory that reproduces the continuum limit of $2$d CDT once it is quantized is indeed 
$2$d projectable HL gravity. The projectable Lapse function allows us to introduce the reparametrization-invariant proper time, 
and to reduce the $2$d field theory to the $1$d system.

\section{Sum over all wormholes and baby universes}
\label{sec:sumoverallgeneraingenera}

In the CDT model, the spatial topology change is not allowed to occur by definition. 
One can generalize the $2$d CDT model in such a way that spatial topology changes do occur in keeping with the foliation structure, 
and the universality class is the same as that of $2$d CDT.   
Such a model is called generalized CDT (GCDT). 
GCDT can be constructed as both discretized and continuum models. 
Here of course the continuum model can be obtained by the continuum limit of the discretized model, 
but one can directly construct the continuum GCDT model promoting the one-dimensional quantum-mechanical system discussed in Sec.~\ref{sec:2dcdt} to a $2$d field theory 
that includes the splitting and joining interactions of the one-dimensional spatial universe. 
Such a field theory is dubbed the string field theory for CDT, in which the string means the one-dimensional universe \cite{Ambjorn:2008ta}. 
In this section, we introduce the string field theory for CDT, and briefly explain the fact that one can take the sum over all wormholes (i.e. handles) and baby universes \cite{Ambjorn:2009wi, Ambjorn:2009fm}. 
Here the baby universe is a portion of geometry that is pinched off from the ``parent universe'' and vanishes into the vacuum. 
We also introduce an effective one-body theory that reproduces the many-body effects coming from the splitting and joining interactions.  
We then discuss those effects in the context of HL gravity \cite{Ambjorn:2021wou}. 
In the end, we show that a sort of Coleman's mechanism works when taking into account all contributions of wormholes and baby universes non-perturbatively \cite{Ambjorn:2021wdm}.

We introduce an operator that creates a marked closed string (i.e. a marked closed one-dimensional universe) with length $L$, $\Psi^{\dagger} (L)$, 
and an operator that annihilates a length-$L$ marked closed string, $\Psi (L)$. 
These operators satisfy the following commutators: 
\[
[\Psi (L), \Psi^{\dagger} (L')] = L \delta (L-L')\ , \ \ \ 
[\Psi (L), \Psi (L')] 
= [\Psi^{\dagger} (L), \Psi^{\dagger} (L')] 
=0\ . 
\label{eq:commutators}
\]   
The vacuum state $| \text{vac} \rangle$ is defined by the equation:  
$\Psi (L) | \text{vac} \rangle  = 0$. 
The CDT amplitude (\ref{eq:matrixelement2}) can be expressed by sandwiching the one-body Hamiltonian: 
\[
\mathcal{G}^{(-1)}_{\Lambda} (L_1,L_2;T) = L_2 G^{(-1)}_{\Lambda} (L_1,L_2;T)
= \langle \text{vac} | \Psi (L_2)\ 
e^{-T \mathcal{H}^{(-1)_{\text{free}}} }\
\Psi^{\dagger} (L_1)
| \text{vac} \rangle\ , 
\label{eq:2ndamplitude-1}
\]
where  
\[
\mathcal{H}^{(-1)}_{\text{free}} 
= \int^{\infty}_{0}  \frac{dL}{L} 
\Psi^{\dagger} (L) \left(
-L \frac{\partial^2}{\partial L^2} + \Lambda L
\right) 
\Psi (L)\ . 
\label{eq:2ndhamiltonian-1}
\]
Hereafter we omit the superscript $(-1)$ for avoiding notational complexity. 
Adding splitting and joining interactions into the free Hamiltonian (\ref{eq:2ndhamiltonian-1}), 
one obtains the full Hamiltonian of the string field theory for CDT:
\[
\mathcal{H}
&=\mathcal{H}_{\text{free}} 
-g_s \int^{\infty}_{0}dL_1 \int^{\infty}_0 dL_2 \Psi^{\dagger} (L_1) \Psi^{\dagger} (L_2)  \Psi (L_1 + L_2) \notag \\ 
&\ \ \ - \alpha g_s \int^{\infty}_{0}dL_1 \int^{\infty}_0 dL_2 \Psi^{\dagger} (L_1 + L_2)  \Psi (L_1)  \Psi (L_2) \notag \\
&\ \ \ - \int^{\infty}_{0} \frac{dL}{L} \delta (L)\Psi (L)\ ,
\label{eq:fullhamiltonian-1}
\]
where the second, third and fourth terms respectively mean the splitting interaction with the string coupling constant $g_s$, 
the joining interaction with the coupling constant $\alpha g_s$, 
and the term associated with a string vanishing into the vacuum. 
Here the parameter $\alpha$ is introduced for counting the number of handles (i.e. wormholes). 
One can in principle calculate the amplitude for the process such that $m$ closed strings propagate in time and end up with 
$n$ closed strings: 
\[
A(L_1, \cdots, L_m; L'_1, \cdots, L'_n; T) 
= \langle \text{vac} | 
\Psi (L'_1) \cdots \Psi (L'_n)
e^{-T \mathcal{H}} 
\Psi^{\dagger} (L_1) \cdots \Psi^{\dagger} (L_m)
| \text{vac} \rangle\ , 
\label{eq:fullamplitude}
\]

\subsection{Effective theory}
\label{sec:effectivetheory}

Let us consider the full propagator $A(L_1 ; L_2;T)$ that includes the sum over all genera and baby universes. 
We can set $\alpha = 1$ without loss of generality since the parameter $\alpha$ plays a supplementary role and we are interested in taking the sum over all genus contributions.  
Somewhat miraculously, the full propagator defined in the many-body system with the Hamiltonian (\ref{eq:fullhamiltonian-1}) can be effectively described by 
the one-body system \cite{Ambjorn:2009wi, Ambjorn:2009fm}: 
\[
A(L_1 ; L_2;T) 
= \langle L_2 | e^{-T \hat{H}_{\text{eff}} (L_1)} | L_1 \rangle\ , 
\label{eq:fullpropagator}
\] 
where $\hat{H}$ is the effective Hamiltonian given by 
\[
\hat{H}_{\text{eff}} (L) = - L \frac{d^2}{dL^2} + \Lambda L - g_s L^2\ . 
\label{eq:effectivehamiltonian}
\] 
This is possible because there exists a bijection called Ambj\o rn-Budd bijection \cite{Ambjorn:2013csx} such that 
one can map each geometry generated in GCDT to a branched polymer with loops at the discrete level. 
The last term in the Hamiltonian (\ref{eq:effectivehamiltonian}), $-g_s L^2$, expresses all the effects originated with the baby universes and the wormholes. 
Note that the Hamiltonian (\ref{eq:effectivehamiltonian}) is not bounded from below because of the last term, 
but in fact this system is known to be ``classical incomplete,'' which means that the Hamiltonian has discrete energy spectra, 
and a set of square integrable eigenfunctions (see e.g. Ref.~\cite{Ambjorn:1992ve} for a pedagogical explanation about the classical incomplete systems). 
A similar deformation has been observed in the $c=1$ non-critical string theory \cite{Moore:1991sf, Betzios:2020nry}. 

The full propagator (\ref{eq:effectivehamiltonian}) can be also described in terms of the path-integral:
\[
A(L_1 ; L_2;T) 
= \int^{L(T) = L_2}_{L(0) = L_1}
\mathcal{D}L(\tau)\ 
\exp \left[
- \int^{T}_{0}d\tau 
\left(
\frac{\dot{L}^2 (\tau)}{4L(\tau)} + \Lambda L(\tau) - g_s L^2 (\tau)
\right)
\right]\ ,
\label{eq:pathintegralfullpropagator}
\]
where the integral measure is given by 
\[
\mathcal{D}L(\tau) 
= \prod^{\tau = T}_{\tau = 0} L^{-1} (\tau) dL (\tau)\ . 
\label{eq:measure-1full}
\]
In order for the functional integral (\ref{eq:pathintegralfullpropagator}) to be well defined, one need to choose the boundary conditions on $L(\tau)$ at infinity such that 
the kinetic term counteracts the unboundedness of the potential. 
If one generalizes the integral measure (\ref{eq:measure-1full}) as 
\[
\mathcal{D}^{(a)}L(\tau) 
= \prod^{\tau = T}_{\tau = 0} L^{a} (\tau) dL (\tau)\ , \ \ \ (a=0, \pm 1)\ ,
\label{eq:measureafull}
\]   
one can recover all possible orderings of the effective Hamiltonian (\ref{eq:effectivehamiltonian}) following the procedure explained in Sec.~\ref{sec:quantization}.

Interestingly, one can reproduce the full propagator (\ref{eq:pathintegralfullpropagator}) 
if one considers that the cosmological constant $\Lambda$ in eq.~(\ref{eq:pathintegral2}) is not a constant but fluctuates independently in time around $\Lambda$, 
following the Gaussian distributions with a standard deviation $\sigma = 2\sqrt{g_s}$:  
\[
A(L_1 ; L_2;T) 
= \int \mathcal{D}\nu (\tau)\ e^{- \frac{1}{4g_s} \int^{T}_{0} d\tau\ \nu^2(\tau)} \mathcal{G}_{\Lambda+\nu} (L_1,L_2;T)\ ,
\label{eq:fluctuations}
\]
where 
\[
\mathcal{G}_{\Lambda+\nu} (L_1,L_2;T) 
:= \int^{L(T)=L_2}_{L(0)=L_1} 
 \mathcal{D}L(\tau)\ 
 \exp \left[ 
 - \int^T_0 d\tau 
 \left(
 \frac{\dot{L}^2(\tau)}{4L(\tau)} + (\Lambda + \nu (\tau)) L(\tau)
 \right)
 \right]\ .
 \label{eq:pathintegralnu}
\]
Therefore, all the contributions coming from the sum over all wormholes and baby universes 
can be fully taken in if the cosmological ``constant'' in (the continuum limit of) $2$d CDT or projectable HL quantum gravity where no wormholes and baby universes exist 
is not really a constant but fluctuates in time. This would lead to a realization of Coleman's mechanism that will be discussed in Sec.~\ref{sec:coleman}.        

In the next section, we will show that the full propagator can be also obtained if quantizing $2$d projectable HL gravity with an effective wormhole interaction term.

\subsection{Wormhole interaction in $2$d projectable HL gravity}
\label{sec:wormholeinteraction}

Let us consider the $2$d projectable HL gravity with a space-like wormhole interaction given by 
the following action: 
\[
I_{\text{w}}  &=
\frac{1}{\kappa} 
\int dtdx\ 
N(t) \sqrt{h(t,x)} 
\left(
(1-\eta) K^2(t,x) - 2 \widetilde{\Lambda}
\right)\notag \\
&\ \ \ + \beta \int dt N(t) \int dx_1 dx_2 \sqrt{h(t,x_1)} \sqrt{h(t,x_2)}
\ , 
\label{eq:wormholeaction}
\]
where $\beta$ is a dimension-full coupling constant. 
The last bi-local term can be interpreted as an effective interaction term for a space-like wormhole connecting two distant regions at a given $t$. 
This bi-local term is allowed to be included since it is invariant under FPD.

Following essentially the same procedure explained in Sec.~\ref{sec:quantization}, 
let us quantize the system defined by the action (\ref{eq:wormholeaction}). 
Introducing the conjugate momentum of the density $\sqrt{h}$ as $\Pi$, we introduce the Poisson bracket (\ref{eq:poisson}). 
Implementing the Legendre transform, one obtains the corresponding Hamiltonian:   
\[
H_{\text{w}} 
= N(t)\ \mathcal{C}_{\text{w}} (t) 
+ \int dx N_1 (t,x)\ \mathcal{C}^1_{\text{w}} (t,x)\ , 
\label{eq:hw}
\]
where 
\[
\mathcal{C}^1_{\text{w}}  (t,x)  
&= - \frac{\partial_1 \Pi (t,x)}{\sqrt{h (t,x)}} \approx 0\ , \label{eq:momentumconst2}\\
\mathcal{C}_{\text{w}}  (t)
&= \int dx\ 
\biggl(
\frac{\kappa}{4(1-\eta)} \Pi^2(t,x) \sqrt{h(t,x)} + \frac{2}{\kappa} \widetilde{\Lambda} \sqrt{h(t,x)} \notag \\
& \ \ \ 
- \beta \sqrt{h(t,x)} 
\int dx_2 \sqrt{h(t,x_2)}
\biggl)
\approx 0 
\ . \label{eq:classicalhamiltonianconst2} 
\] 
The constraints (\ref{eq:momentumconst2}) and (\ref{eq:classicalhamiltonianconst2}) are the momentum constraint and the Hamiltonian constraint, respectively. 
Solving the momentum constraint (\ref{eq:momentumconst2}) at the classical level as before, 
the Hamiltonian (\ref{eq:hw}) reduces to the following one-dimensional one:
\[
H_{\text{w}}
= N(t) \left(
\frac{\kappa}{4(1-\eta)} \Pi^2 (t) L(t) 
+ \frac{2}{\kappa} \widetilde{\Lambda} L(t) 
- \beta L^2 (t)
\right)\ , 
\label{eq:hw2}
\] 
where $L(t) := \int dx \sqrt{h(t,x)}$. The Hamiltonian (\ref{eq:hw2}) is subject to the Hamiltonian constraint: 
\[
L(t)
\left(
\frac{\kappa}{4(1-\eta)} \Pi^2 (t) 
+ \frac{2}{\kappa} \widetilde{\Lambda} 
- \beta L (t) 
\right)
\approx 0 
\ .  
\label{eq:classicalhamiltonianconst3}
\]  
Here we choose the CDT parametrization (\ref{eq:parameteridentification}). 
A solution to the Hamiltonian constraint (\ref{eq:classicalhamiltonianconst3}) is 
\[
\Pi^2 = - \Lambda + \beta L \ge 0\ , \ \ \ \text{for}\ \ \ \sqrt{\Lambda} L \ge 1/\xi\ , 
\label{eq:solutiontohamiltonianconstw1}
\]  
where $\xi$ is a dimensionless parameter given by $\xi = \beta/ \Lambda^{3/2}$. 
For $\sqrt{\Lambda} L < 1/\xi$, the only allowed solution is $L=0$.

When quantizing the system, if we follow the same procedure described in Sec.~\ref{sec:quantization}, and set 
$\beta = g_s$, one can reproduce the path-integral of the full propagator (\ref{eq:pathintegralfullpropagator}). 
Remember the boundary condition for the path-integral (\ref{eq:pathintegralfullpropagator}), i.e. 
the kinetic term should counteract the unboundedness of the potential term at $L=\infty$. 
This balance between the kinetic and potential terms is precisely what is reflected in the classical Hamiltonian constraint (\ref{eq:solutiontohamiltonianconstw1}).

\subsection{Coleman's mechanism} 
\label{sec:coleman}

In this section, we discuss a sort of Coleman's mechanism in the context of two-dimensional gravity based on CDT briefly.  

Let us define the two kinds of Wheeler-deWitt equation: 
\[
\hat{H} W_0 (L) = 0\ , \ \ \ \hat{H}_{\text{eff}} W(L) = 0\ , 
\label{eq:wdw}
\]
where $\hat{H} := \hat{H}^{(-1)}$ introduced in eq.~(\ref{eq:hamiltonians}). 
The solutions to the Wheeler-deWitt equations are the Hartle-Hawking wave functions given by  
\[
W_0 (L) = e^{-\sqrt{\Lambda} L}\ , \ \ \ 
W(L) = \frac{\text{Bi} (\xi^{-2/3} - \xi^{1/3} \sqrt{\Lambda} L)}{ \text{Bi} (\xi^{-2/3}) } 
+ c\ \text{Ai} ( \xi^{-2/3} - \xi^{1/3} \sqrt{\Lambda} L)\ , 
\label{eq:hartlehawking}
\]    
where $\text{Ai}$ and $\text{Bi}$ are the standard Airy functions, 
$\xi$ is a dimensionless string coupling constant measured by the cosmological constant, i.e.,  
$\xi := g_s/ \Lambda^{3/2}$;  
and $c$ is an undetermined dimensionless constant. 
The Hartle-Hawking wave function $W_0 (L)$ is the one for (the continuum theory of) $2$d CDT, 
i.e. neither baby universe nor wormhole contributions are included. 
On the other hand, $W(L)$ is the Hartle-Hawking wave function including all possible contributions of baby universes and wormholes non-perturbatively.      

\begin{figure}[h]
\centering
\includegraphics[scale=.4]{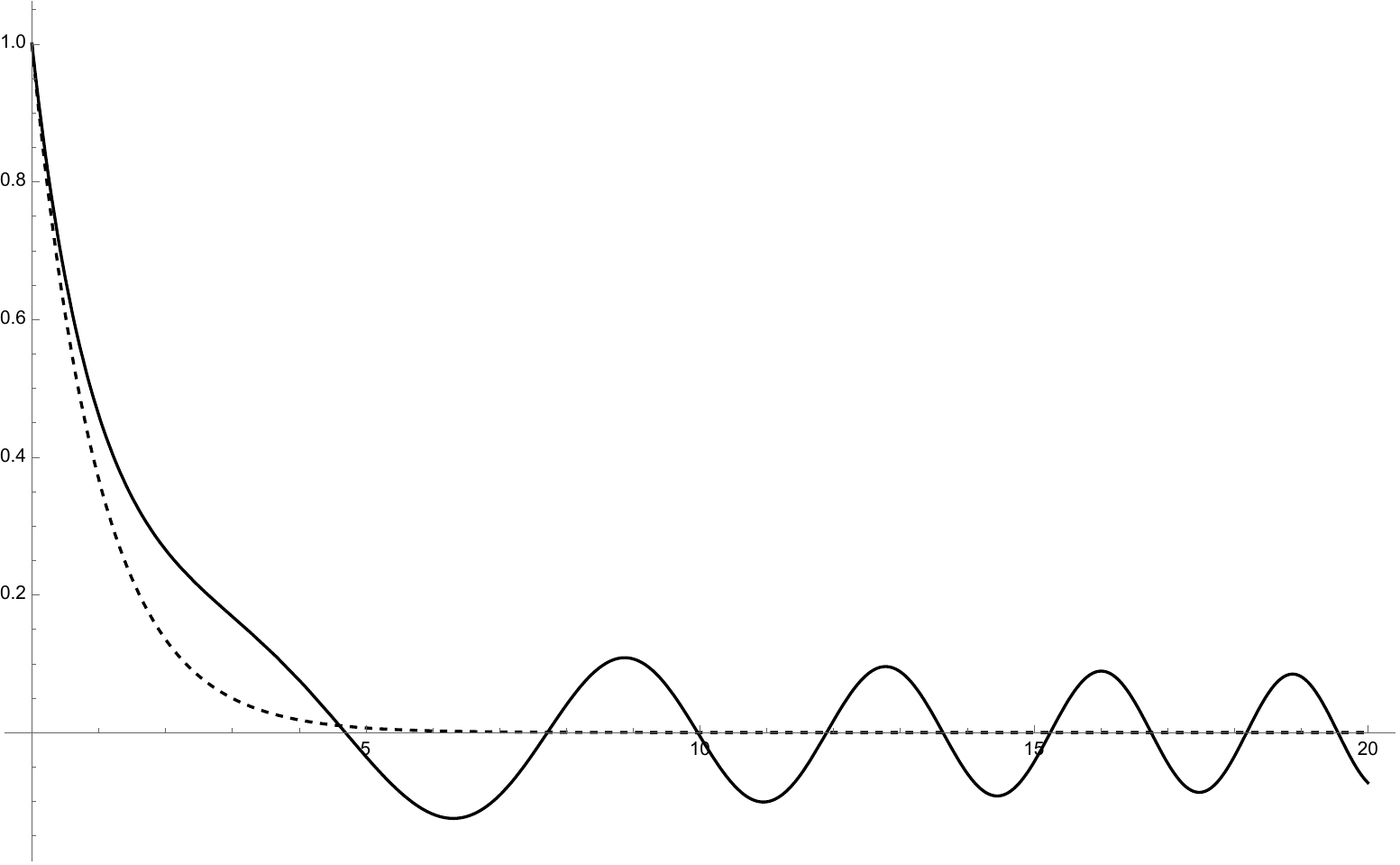}
\caption{A plot of the Hartle-Hawking wave functions, $W_0$ (dashed line) and $W$ (solid line), for $\xi=1/3$ and $c=0$:  
The horizontal axis is $\sqrt{\Lambda} L$ and the vertical axis is either $W_0$ or $W$.  
}
\label{fig:HH}       
\end{figure}

We wish to explore the behavior of the non-perturbative Hartle-Hawking wave function $W(L)$ (see Fig.~\ref{fig:HH}).
For $\sqrt{\Lambda}L \ll 1/\xi$, one obtains the asymptotic expansion: 
\[
W(L) \sim e^{-\sqrt{\Lambda}L} = W_0(L)\ . 
\label{eq:asymptoticexpansion}
\] 
Therefore, when the size of the one-dimensional universe is small enough, 
the physics is very close to the one without baby universes and wormholes, and it is essentially governed by the cosmological constant.  
The wave function in this region decreases exponentially, and this behavior does not change at any finite order of perturbation.   
 
However, once the size of the universe is large enough, i.e. $\sqrt{\Lambda}L > 1/\xi$, the wave function starts oscillating, and the behavior is governed by 
the string coupling constant instead of the cosmological constant:
\[
W(L) \sim 1/(g^{1/3}_s L)^{1/4}\ . 
\label{eq:largebehavior}
\] 
This drastic change happens due to the infinitely many wormholes and baby universes. 
The similar behavior has been observed in the context of non-critical string theory \cite{Moore:1991sf, Betzios:2020nry}. 

From the discussion above, we observe that a sort of Coleman's mechanism works:  
For a large universe, the cosmological constant is not important enough to govern the physics\footnote{
The Hartle-Hawking wave function $W(L)$ is not normalizable, but similar arguments are valid on the normalizable energy eigenstates \cite{Ambjorn:2021wdm}. 
}.

\section{Summary}
\label{sec:summary} 

We have reviewed the relation between two-dimensional causal dynamical triangulations ($2$d CDT) and two-dimensional projectable Ho\v rava-Lifshitz quantum gravity ($2$d projectable HL QG). 

In the first part, it has been shown that the physics described by the continuum limit of $2$d CDT coincides with the one obtained quantizing $2$d projectable HL gravity. 
This is confirmed because the quantum Hamiltonians of both models are exactly the same. 
The system is expressed in terms of quantum mechanics of a $1$d extended object, i.e. a $1$d universe.   

It would be too hasty to consider that this scenario also holds for the higher dimensional cases. 
In fact, it has been shown that in $2+1$ dimensions numerical studies of the so-called locally causal dynamical triangulations that relax the proper time foliation of CDT and require the local causality reproduce an intriguing specialty of CDT, 
an emergence of the de-Sitter-like geometry \cite{Jordan:2013awa} (see e.g. Ref.~\cite{Loll:2019rdj} for the higher-dimensional CDT). 
On the other hand, a Landau theory approach suggests a relation between CDT and theories invariant under the foliation-preserving diffeomorphisms in $2+1$ dimensions \cite{Benedetti:2014dra,Benedetti:2016rwo} 
(see also the dedicated chapter of the Handbook of Quantum Gravity \cite{Benedetti:2022ots}). 
This issue should be investigated further.          

In the second part, we have introduced the generalized CDT (GCDT) that permits baby universes and wormholes to form in keeping with the foliation, 
and in particular the construction based on the string field theory for CDT has been explained. 
Here the string means the $1$d universe, and the string field theory is constructed in such a way that the free part reproduces the CDT amplitudes, 
and the splitting and joining interactions of string are introduced to create baby universes and wormholes.

Focusing on the loop-to-loop amplitude, we have introduced an effective $1$d theory that includes all the contributions coming from the sum over all possible baby universes and wormholes. 
From the point of view of HL gravity, the effective theory can be precisely reproduced if introducing a bi-local interaction term into the action of $2$d projectable HL gravity and if quantizing the system. 
In addition, the effective theory can be also obtained considering that the cosmological constant of $2$d CDT is not a constant but it fluctuates in time. 
This leads to Coleman's mechanism in $2$d CDT such that for a large universe the cosmological constant is not important enough to govern the physics.

Although we have not discussed issues of the coupling to matter, $2$d CDT coupled to Yang-Mills theory has been solved analytically in Ref.~\cite{Ambjorn:2013rma}, 
and it has been shown that the quantum Hamiltonian obtained in Ref.~\cite{Ambjorn:2013rma} can be reproduced quantizing $2$d projectable HL gravity coupled to Yang-Mills theory \cite{Ipsen:2015ckl}. 
In fact, we know very little about the analytical treatment of the coupling to matter compared to the situation of $2$d dynamical triangulations and the Liouville quantum gravity. 
This direction needs to be explored in the future.

What is remarkable is that following the standard Wilsonian renormalization group, one can take the continuum limit of the lattice model, $2$d CDT, 
and find the continuum quantum field theory, $2$d projectable HL QG, which is in the same universality class of $2$d CDT. 
A missing piece is the continuum quantum field theory of GCDT that is described by metric components and allows us to compute all the amplitudes defined by the string field theory for CDT, 
although we have the effective field theory, the $2$d projectable HL gravity with a bi-local interaction, that reproduces the restricted class of GCDT amplitudes once it is quantized. 
We wish to unveil the underlying continuum quantum field theory of GCDT, through which we can understand something inherently interesting about quantum geometries for sure.

\begin{acknowledgement}
YS would like to thank 
Jan Ambj\o{}rn, 
Lisa Glaser, 
Yuki Hiraga, 
Yoshiyasu Ito 
and 
Yoshiyuki Watabiki, for wonderful collaborations on the topics related to this chapter. 
Great ideas appeared in this chapter attribute to them. However, if conceptual mistakes exist, YS is responsible for them. 
This work was partially supported by JSPS KAKENHI Grant Number 19K14705.    
\end{acknowledgement}

\end{document}